\pdfoutput=1
\documentclass{article}

\usepackage[symbol]{footmisc}
\usepackage{amssymb,amsfonts,amsmath,stmaryrd}
\usepackage{color}

\def\be{\begin{eqnarray}}
\def\ee{\end{eqnarray}}
\def\nn{\nonumber}

\def\p{\partial}
\def\tr{{\rm tr}\,}
\def\Tr{{\rm Tr}\,}

\newcommand{\beq}{\begin{equation}}
\newcommand{\eeq}{\end{equation}}
\newcommand{\beqa}{\begin{eqnarray}}
\newcommand{\eeqa}{\end{eqnarray}}

\definecolor{red}{rgb}{1,0,0}
\definecolor{orange}{rgb}{1,0.5,0}
\definecolor{violet}{rgb}{0.7,0,1}

\hoffset= - 1.0in         

\begin{document}

\begin{center}
\begin{small}
FIAN/TD-01/21 \hfill\phantom.\\
IITP/TH-01/21 \hfill\phantom.\\
ITEP/TH-01/21 \hfill\phantom.\\
MIPT/TH-01/21 \hfill\phantom.\\
\end{small}

\vspace{-0.9cm}

{\bf
\hfill to the memory of\\
\hfill  Sergey Natanzon\\
}
\end{center}

\vspace{1cm}

\begin{center}

\begin{Large}\fontfamily{cmss}
\fontsize{17pt}{27pt}
\selectfont
	\textbf{Generalized Q-functions for GKM }
	\end{Large}
	
\bigskip \bigskip
\begin{large}A. D. Mironov$^{a,b,c}$\footnote{mironov@lpi.ru; mironov@itep.ru},
A. Morozov$^{d,b,c}$\footnote{morozov@itep.ru}
 \end{large}
\\
\bigskip

\begin{small}
$^a$ {\it Lebedev Physics Institute, Moscow 119991, Russia}\\
$^b$ {\it ITEP, Moscow 117218, Russia}\\
$^c$ {\it Institute for Information Transmission Problems, Moscow 127994, Russia}\\
$^d$ {\it MIPT, Dolgoprudny, 141701, Russia}\\
\end{small}
 \end{center}
\medskip

\begin{abstract}
Recently we explained that the classical $Q$ Schur functions
stand behind various well known properties of the cubic Kontsevich model,
and the next step is to ask what happens in this approach to
the generalized Kontsevich model (GKM)
with monomial potential $X^{n+1}$.
We suggest to use the Hall-Littlewood polynomials at the parameter equal to the $n$-th root of unity
as a generalization of the $Q$ Schur functions from $n=2$ to arbitrary $n>2$.
They are associated with $n$-strict Young diagrams and
are independent of time-variables $p_{kn}$ with numbers divisible by $n$.
These are exactly the properties possessed by the generalized Kontsevich model (GKM),
thus its partition function can be expanded in such functions $Q^{(n)}$.
However, the coefficients of this expansion remain to be properly identified.
At this moment, we have not found any ``superintegrability" property
$<character>\,\sim character$, which expressed these coefficients through
the values of $Q$ at delta-loci in the $n=2$ case.
This is not a big surprise, because for $n>2$ our suggested $Q$ functions
are not looking associated with characters.
\end{abstract}

\bigskip

\section{Introduction}

Serezha Natanzon was a very original scientist.
He taught us a lot about his beloved Hurwitz numbers
and related world of interesting special functions.
One of these lessons was about the $Q$ Schur functions,
which, in his opinion, had to be applied to
counting of holomorphic coverings with spin structures
(spin Hurwitz numbers), and they really did \cite{MMNspin}.
Once one knows about the $Q$ Schur functions, it gets immediately clear
that they are applicable to the Kontsevich model \cite{Kon},
because they have just the needed properties:
do not depend on even times, are labeled by special Young
diagrams, satisfy the BKP equations, and behave nicely
under action of the Virasoro algebra \cite{VirJ}.
Since the $Q$ Schur functions are related to characters
(of the Sergeev group \cite{Serg,Sergrev}),
it comes without a surprise that they exactly what is necessary
to formulate the {\it superintegrability} property
\cite{super,MMcharchar}
$<character>\,\sim character$ of the Kontsevich model
\cite{DIZ,MMkon}, moreover, as usual in this case,
the coefficients at the r.h.s are made from the same
$Q$ Schur functions at special delta-loci.
At least in this case, this follows from magnificent
factorization formulas for the $Q$ Schur characters \cite{Alex,MMNO}.

Knowing all this, one is tempted to move towards the
generalized Kontsevich model \cite{GKM,UFN3} (GKM),
with a monomial potential $X^{n+1}$ to begin with.
The first obvious question is what are the relevant
$Q^{(n)}$ functions, which were just Schur's $Q=Q^{(2)}$
for $n=2$.
In this paper, we suggest as a plausible candidate the
Hall-Littlewood polynomials at the $n$-th root of unity, which is a straightforward generalization of {\it one of}
the many definitions at $n=2$.
We explain that at least some of the needed properties
are captured by this suggestion.
But some are not, and Serezha is no longer here to
teach us how to resolve these problems.
They remain for the future work.

\section{Generalized $Q$ Schur polynomials}

In this section, we briefly list the main properties of
a generalization of $Q$ Schur polynomials from $n=2$ to $n>2$.
To this end, we need a couple of new quantities:
for the Young diagram $\Delta=\{\delta_1\geq \delta_2\geq\ldots\delta_{l_\Delta}>0\}
= \{1^{m_1},2^{m_2},\ldots\}$, we define
the standard symmetry factor
\be
z_\Delta:=\prod_{a=1}^\infty m_a! \cdot a^{m_a}
\ee
and two less conventional $n$-dependent quantities
\be
L_\Delta:=\prod_{a=1}^\infty \prod_{k=1}^{m_a}\beta_k
\ee
and
\be
\beta_\Delta:=\prod_i^{l_\Delta}\beta_{\delta_i}
\ee
where
\be
\beta_k^{(n)}:=1-\omega^k_n = 1 - e^{\frac{2\pi ik}{n}}
\ee
Thus $L_\Delta$ is vanishing when diagram $\Delta$ has $n$ (or more)
lines of the same length.
Roots of unity of the degree dictated by power of the monomial potential in the GKM
is a new twist in the QFT-Galois relation with far-going applications.

\bigskip

The properties of our new functions $Q_R^{(n)}\{p\}$ are:

\begin{itemize}

\item
They are equal to the special values of the Hall-Littlewood polynomials \cite{Mac}:
at $t=\omega_{n}$ which is the primary $n$-th root of unity
\be
Q^{(n)}_R:=
\sqrt{L_R}\cdot\hbox{Mac}_R(q=0, t = e^{2i\pi/n})
\ee
where $\hbox{Mac}_R$ denotes the Macdonald polynomial.
Because of the factor $L_R$, $Q^{(n)}_R$
is non-zero only for the Young diagram $R$
which has no more than $n-1$ lines of the same length.
We denote this set of diagrams by $S_nP$, and its subset of diagrams of size $|R|=m$, by $S_nP(m)$.

\item Hereafter we deal with the symmetric functions $Q^{(n)}_R$ of variables $x_i$ as (graded) polynomials of power sums $p_k:=\sum_ix_i^k$, which we call time-variables (they are proportional to times of integrable hierarchy associated with the GKM). The polynomials $Q^{(n)}_R\{p\}$ are independent of the time-variables $p_{kn}$.
This property is preserved by an arbitrary rescaling of time-variables,
but we do not apply this rescaling, and use the choice associated with the Cauchy formula in the form
(\ref{QCauchy}) below.

\item{They form a closed algebra }
\be
Q^{(n)}_{R_1}\{p\}\cdot Q^{(n)}_{R_2}\{p\} = \!\!\!\sum_{{R \in R_1\otimes R_2}\atop{R\in {\tiny S_nP}}}\!\!\!
N_{R_1,R_2}^R Q^{(n)}_R\{p\}
\ee
i.e. the Littlewood-Richardson coefficients of Macdonald polynomials
vanish when $q=0, t=\omega_{n}$ and $R_1,R_2\in S_nP$ and $R\notin S_nP$.
Like in the case of $n=2$,
$\hbox{Mac}_R(q=0,t=\omega_{n})$ themselves do not vanish for $R\notin S_nP$,
and then they can also depend on even $p_{nk}$,
thus the set of $Q^{(n)}_R\{p\}$ is a {\it sub-set}
of that of the Hall-Littlewood polynomials, and
it is a non-trivial fact that it is a {\it sub-ring}.

\item
The Fr\"obenius formula for the generalized
$Q$ Schur polynomials is
\be\label{FQ}
Q_R^{(n)}\{p_k\}=\sum_{\Delta\in  O_nP} {\Psi^{(n)}_R(\Delta)\over z_\Delta}p_\Delta
\ee
At the l.h.s. $R\in S_nP$.
The set $O_nP$ at the r.h.s. consists of all diagrams with line lengths not divisible by $n$,
and this formula reflects  a remarkable one-to-one correspondence between $O_nP$ and $S_nP$: these sets have the same sizes, and the map is non-degenerate as follows from the orthogonality relations:
\be\label{QOR}
\sum_{\Delta\in {\footnotesize O_nP}}{\Psi_R^{(n)}(\Delta)\Psi_{R'}^{(n)}(\Delta)\over \beta_\Delta z_\Delta}=\delta_{RR'},
\ \ \ \ \ \ \
\sum_{R\in {\footnotesize S_nP}}{\Psi_R^{(n)}(\Delta)\Psi_{R}^{(n)}(\Delta') }=
\beta_\Delta z_\Delta
\delta_{\Delta\Delta'}
\ee
Hence, one can construct an inverse map
\be\label{FQi}
p_\Delta =\sum_{R\in  S_nP}{\Psi^{(n)}_R(\Delta)Q_R^{(n)}\{p_k\}\over  \beta_\Delta}
\ee

\item In the scalar product
\be
\Big< p_{k} \Big| p_{l}\Big> = {k\over\beta_k} \cdot \delta_{k,l}
\ee
the $Q$-functions are orthogonal:
\be
\Big< Q_R^{(n)}\Big| Q_{R'}^{(n)} \Big> = ||Q_R^{(n)}||^2\cdot\delta_{R,R'}
\ee
with
\be
||  Q_R^{(n)}||^2 = 1
\ee

\item As usual, one can introduce the skew $Q$-Schur functions $Q^{(n)}_{R/P}$ defined as
\be
Q^{(n)}_{R}\{p+p'\}=\sum_{P\in\hbox{\footnotesize SP}}Q^{(n)}_{R/P}\{p\}Q^{(n)}_P\{p'\}
\ee
They are given by
\be
Q^{(n)}_{R/P}\{p\}=\sum_{P\in\hbox{\footnotesize SP}}{\cal N}_{PS}^{R}Q^{(n)}_S\{p\}
\ee

\item
$Q$ Schur polynomials satisfy the Cauchy formula,
\be\label{QCauchy}
\sum_{R\in {\footnotesize S_nP}} Q^{(n)}_{R}\{p\}\cdot Q^{(n)}_{R}\{\Tr X^k\}
= \exp\left(\sum_{k=1} \frac{\beta_k^{(n)}\, p_{k}\, \Tr X^{k}}{k}\right)
\ee
Since $\beta_k^{(n)}$ vanishes whenever $k$ is divisible by $n$,
the r.h.s. is independent of all $p_{kn}$.
We write the second set of times in Miwa variables, $p_k'=\Tr X^k$,
because we need this form of the Cauchy formula in consideration of correlators below.

\item The Virasoro and $W$ algebras act rather simple on the generalized $Q$ Schur polynomials, see sec.\ref{VW}.

\item The $Q$ polynomials themselves are {\it not} $\tau$-functions of
the KP hierarchy and its reductions (like the KdV and Boussinesq ones).
Instead, for $n=2$ they satisfy the BKP hierarchy \cite{You,O2003,MMNspin} which does not yet have any
direct counterpart for $n>2$.

\item The main difficulty at this stage is that there is yet no formula
for $Q_R^{(n)}$ per se, without referring to the Hall-Littlewood and Macdonald polynomials.
Indeed, the ordinary Schur polynomials at $n=1$ have a determinant representation \cite{Mac}, or can be realized as an average over charged fermions (see a review in \cite{JM}); the $Q$ Schur polynomials at $n=2$ have a Pfaffian representation instead of the determinant one (see, e.g., \cite[Eq.(74)]{MMNspin}), or can be realized as an average over neutral fermions \cite{DJKM,JM,You,O2003,MMNO};
what happens for $n>2$ is yet unclear.
One could expect some expressions in terms of parafermions,
which generalizes the reduction from charged to neutral fermions in the $n=2$ case.
\end{itemize}

\section{Monomial GKM
\label{GKMdef}}

\subsection{Properties of GKM \cite{GKM,UFN3}}

The monomial Generalized Kontsevich model is defined by the $N\times N$ Hermitian matrix integral \cite{GKM}
\be\label{GKMint}
Z^{(n)}(L) :={\cal N}(L)\cdot \int  \exp\left(- {\Tr X^{n+1}\over n+1}+\Tr L^{n} X\right) dX
\ee
$Z^{(n)}(L)$ depends only on the eigenvalues of the background matrix field $L$,
and, with a proper choice of the normalization factor ${\cal N}(L)$,
it can be treated as a formal series either in positive or in negative powers of $L$ \cite{GKMU}.
In fact, $Z^{(n)}(L)$ is a {\it symmetric} functions of the eigenvalues $\lambda_i^{\pm 1}$ of the external matrix $L$,
and, hence, can be considered as a function of the power sums or the ``time-variables"  $p_k^{\pm} := \tr L^{\pm k}$.
These two cases require proper (different) choices of the normalization factors and are referred to as {\it character} and {\it Kontsevich} phases \cite{GKMU}. In this paper, we are interested in the more sophisticated Kontsevich phase,
and in what follows we omit the superscript "-":  $p_k:=p^-_k$.

The potential in the exponent has an extremum at $X=L$,
and, in the Kontsevich phase, one expands around it
in inverse power of $L$. In this phase, one has to choose the normalization factor
\be
{\cal N}(L):=
{\displaystyle{\exp\left({1\over n+1}\Tr L^{n+1}\right)}\over
\displaystyle{\exp\left(-\frac{1}{2}\sum_{a+b=n-1} \Tr L^a X L^b X\right) dX}}
\ee
This provides that $Z^{(n)}(L)$, which depends on the eigenvalues $\lambda_i$ of the matrix $L$, can be understood as a formal power series in $\lambda_i^{-1}$, and, in fact, is a power series in $p_k:=\Tr L^{-k}$ \cite{GKM}.

It possesses more advanced definitions as a $D$-module and/or peculiarly reduced
KP $\tau$-function.
Namely,

\begin{itemize}

\item{
For a given $n$ the partition function $Z^{(n)}(L)$ is actually independent of $p_{kn}$, this explains the choice of notation for
the potential: what matters is usually not the potential $X^{n+1}$ but its
derivative $X^n$.
}

\item{
$Z^{(n)}(L)$ as a (symmetric) function of $\lambda_i$ is a $\tau$-function of the KP hierarchy in Miwa variables,
i.e. satisfies the bilinear difference Hirota equations
and can be expressed as a determinant. $Z^{(n)}(L)$ as a function of power sums $p_k/k$ is a $\tau$-function of the KP hierarchy in the ordinary higher time variables (hence, the name ``time-variables" for $p_k$), and satisfies the bilinear differential Hirota equations.
Moreover, for a given $n$, it is actually an $n$-{\it reduction} of the KP hierarchy,
say, the KdV hierarchy for $n=2$, or the Boussinesq hierarchy for $n=3$.
}

\item{
$Z^{(n)}(L)$ satisfies the Ward identities \cite{MMM91,GN}.
When rewritten in terms of $p^-_k$, these constraints form Borel subalgebras of the Virasoro and $W$-algebras, $\hat W^{(p)}_mZ (L)= 0$
with $2\leq p\leq n$, $m\ge 1-p$ \cite{GKM,Mikh}.
In the character phase, the Ward identities in terms of $p^+_k$
are rather the $\tilde W$-constraints \cite{GKMU}.
}

\item{The lowest of these constraints, $\hat L_{-1}Z^{(n)}(L)=\hat W^{(2)}_{-1}Z^{(n)}(L)=0$ called string equation along with the integrable hierarchy equations generates the whole set of the Ward identities.
}

\end{itemize}

We see that this list has some parallels with the list of properties
of the $Q$-functions in the previous section.
Thus it comes without a surprise that

\begin{itemize}

\item{$Z^{(n)}(L)$ in the Kontsevich phase can be expanded in functions $Q^{(n)}\{p\}$.}

\end{itemize}

This {\it character expansion} is the subject of the present paper.
We will see that it is not yet as powerful as in the case of $n=2$ \cite{MMkon},
still generalization to $n>2$ clearly exists.

\subsection{The GKM propagator}

One can calculate the GKM integral (\ref{GKMint}) perturbatively. To this end, one has to expand around the extremum of potential at $X=L$, i.e. to shift $X=L+Y$, and deal with the integral
\be
Z^{(n)}(L)= \nn\\
=\int \exp \left(-\Tr \frac{(L+Y)^{n+1} - L^{n+1}}{n+1} +L^{n}Y +\frac{1}{2}\sum_{a+b=n-1} \Tr L^a Y L^b Y\right) \exp\left(-\frac{1}{2}\sum_{a+b=n-1} \Tr L^a Y L^b Y\right) dY
\label{GKMkph}
\ee
expanding the first exponential and evaluating the obtained Gaussian integral. The measure in this integral is defined so that $<1>=1$.

Thus, we define the correlation function by the Gaussian integral
\be
\left< \ldots \right> \ :=
\int \ldots \exp\left(-\frac{1}{2}\sum_{a+b=n-1} \Tr L^a Y L^b Y\right) dY
\ee
and first evaluate the propagator.
In terms of the eigenvalues $\lambda_i$ of $L$,
the propagator is
\be
\left<Y_{ij}Y_{kl}\right>_n=\frac{ \delta_{il}\delta_{jk}}
{\!\!\!\!\displaystyle{\sum_{a+b=n-1}}  \lambda_i^a \lambda_j^b\ }
\label{prop}
\ee
When this does not lead to a confusion, in what follows we omit the index $n$ in the notation of the average,
but we should remember that the propagator depends on $n$ and has grading level, i.e. the power in $L^{-1}$
equal to $n-1$.

\subsection{Correlation functions}

Correlation functions with the propagator (\ref{prop}) have complicated denominators
and often can not be expressed in terms of the time-variables
\be
p_k=\tr L^{-k} = \sum_{i=1}^N \lambda_i^{-k}
\ee
From this perspective it looks like a miracle that there are many exceptions:
plenty of {\it admissible} correlators exist, i.e. those expressible through the time-variables.
In particular, as we already pointed out an important result from the theory of GKM \cite{GKM}
is that $Z^{(n)}(L)$ in the Kontsevich phase
actually {\it is} a power series in time variables.

Thus we understand that at least the correlators
which comes from perturbative expansion of the GKM are admissible.
In fact, expanding exponential in (\ref{GKMkph}),
one obtains rather sophisticated averages
\be
\left< \left(\Tr \frac{(L+Y)^{n+1} - L^{n+1}}{n+1} - L^{n}Y -\frac{1}{2}\sum_{a+b=n-1} \Tr L^a Y L^b Y\right)^m
\right>_{\! n}
\ee
and they should depend on $\lambda_i$ only through $p_k$.
For the ordinary Kontsevich model with $n+1=3$, these  correlators are just  $\left< \left(\Tr Y^3\right)^{2m} \right>$,
but already in the quartic case, $n+1=4$ they contain $L$: powers of $ \Tr Y^4$
should be combined with those of $\Tr LY^3$.

These correlators do not exhaust all the admissible correlators, but, as a first step, we concentrate on this special
set in this paper.

\subsection{Character expansion}

Now we are going to study the perturbative expansion of (\ref{GKMkph}) as a function of time-variables.
A natural full basis in the set of such functions is provided by {\it characters},
for instance, by the Schur functions $\chi_R\{p\}$, which form a set labeled by the Young diagrams with a natural grading by the size of these diagrams $|R|$:
\be
Z^{(n)}(L) =  \sum_R C_R\chi_R\{p\}
\ee
The question is what are the coefficients $C_R$.
Since $Z^{(n)}(L)$ is a KP $\tau$-function, they satisfy the Pl\"ucker relations.
But actually from the theory of GKM \cite{GKM} we know more:
for a given $n$, it is {\it independent} of all $p_{kn}$
and is a $\tau$-functions of the (appropriately reduced) KP hierarchy.
For example, at $n+1=3$, it is a $\tau$-function of the KdV hierarchy,
which depends only on odd time-variables $p_{2k+1}$.
This means that $\chi_R\{p\}$ is actually not the most adequate basis,
because this type of reduction looks complicated in it. This is clear already from the case of $n=2$, where the coefficients $C_R$ are quite involved \cite{Zh,BY}.

From what we already know from sec.2, it is clear that much better for a given $n$
is a basis formed by the $Q$-functions $Q^{(n)}_R$ with $R\in S_nP$.
Thus, more precisely, our interest is in
\be
\boxed{
Z^{(n)}(L) = \sum_{R \in S_nP}  C^{(n)}_R \,Q_R^{(n)}\{p\}
}
\label{GKMQexpan}
\ee
As we demonstrated in \cite{MMkon}, the coefficients $C_R^{(2)}$ in this basis are very simple and natural in the case of $n=2$, in contrast with expansion into $C_R$.

\paragraph{Remark.}
It could look appealing to extract
$Q^{(n)}$ at a special delta-locus, $Q_R^{(n)}\{\delta_{k,n+1}\}$
from the coefficients $C_R^{(n)}$:
\be
Z^{(n)}(L) \stackrel{?}{=} \sum_{R \in S_nP}  c^{(n)}_R\cdot Q_R^{(n)}\{\delta_{k,n+1}\}\,Q_R^{(n)}\{p\}
\label{Znc}
\ee
like we did in \cite{MMkon} for $n=2$.
This may seem natural because applying the Cauchy identity to the original integral (\ref{GKMint}),
one can conclude that
\be
e^{-\frac{1}{n+1}\,\Tr X^{n+1}} = \exp\left(-\sum_k \frac{1}{k}\,\Tr X^k \cdot \delta_{k,n+1}\right) = \sum_R (-1)^{|R|}\,Q^{(n)}_{R^\vee}\{\Tr X^k\} \cdot Q^{(n)}_R\{\delta_{k,n+1}\}
\ee
Of course, this is far from a reliable argument,
and it is not a big surprise that things are not so simple.
As we will see shortly, the expansion (\ref{Znc}) is actually not possible for $n>2$.

\subsection{The reminder from \cite{MMkon}: $n=2$}

For $n+1=3$ the relevant averages involve only $L$-independent operators:
$\sum_{m=0}^\infty \frac{1}{(2m)!\cdot 3^{2m}}\left< (\Tr Y^3)^{2m} \right>$.
For example, for the first two terms
\be
\frac{1}{2!\cdot 3^2}\left< (\Tr Y^3)^{2} \right> = \frac{1}{48}\cdot\left(p_3+4p_1^3\right) =
\frac{1}{48}\cdot\left(Q^{(2)}_{[2,1]}\{p_k\}-\frac{5\sqrt{2}}{2}\,Q^{(2)}_{[3]}\{p_k\} \right)
=\nn \\
= \frac{1}{32}\cdot\left(5 \,Q^{(2)}_{[3]}\{\delta_{k,3}\}Q^{(2)}_{[3]}\{p_k\} - Q^{(2)}_{[2,1]}\{\delta_{k,3}\}Q^{(2)}_{[2,1]}\{p_k\}\right)
\ee
and
\be
\frac{1}{4!\cdot 3^4}\left< (\Tr Y^3)^{4} \right>
= \frac{1}{9\cdot 512}\cdot\left(144p_5p_1 + 25p_3^2+200p_3p_1^3+16p_1^6\right)
=\ \ \ \ \ \nn \\
=- \frac{5}{9\cdot 512}\cdot\left( \frac{77\sqrt{2}}{2}\,Q^{(2)}_{[6]}\{p_k\}
- 7\,Q^{(2)}_{[5,1]}\{p_k\} + 7\,Q^{(2)}_{[4,2]}\{p_k\} + \sqrt{[2]}Q^{(2)}_{[1,2,3]}\{p_k\}
\right) =
\ee
\vspace{-0.4cm}
\be
= \frac{5}{1024}\cdot\left( 77 \,Q^{(2)}_{[6]}\{\delta_{k,3}\}Q^{(2)}_{[6]}\{p_k\}
- 7\,Q^{(2)}_{[5,1]}\{\delta_{k,3}\}Q^{(2)}_{[5,1]}\{p_k\} - 7\,Q^{(2)}_{[4,2]}\{\delta_{k,3}\}Q^{(2)}_{[4,2]}\{p_k\} - Q^{(2)}_{[1,2,3]}\{\delta_{k,3}\}Q^{(2)}_{[1,2,3]}\{p_k\}
\right)
\nn
\ee
We can note that the coefficient in front of  $Q^{(2)}_{[1,2,3]}\{\delta_{k,3}\}Q^{(2)}_{[1,2,3]}$
is exactly the product of those in front of $Q^{(2)}_{[3]}\{\delta_{k,3}\}Q^{(2)}_{[3]}\{p_k\}$
and $Q^{(2)}_{[2,1]}\{\delta_{k,3}\}Q^{(2)}_{[2,1]}\{p_k\}$,
i.e. $\,-\frac{ 5}{1024} = \frac{ 5}{32} \cdot \left(-\frac{1}{32}\right)$.
This is a manifestation of the general property \cite{Alex} that the coefficient
in front of $Q^{(2)}_{R}\{\delta_{k,3}\}Q^{(2)}_{R}\{p_k\}$ in the character expansion
of the cubic Kontsevich partition function $Z^{(2)}(L)$ is factorized: equal to
\be
{\rm coeff}_{Q^{(2)}_{R}\{\delta_{k,3}\}Q^{(2)}_{R}\{p_k\}}\Big(Z^{(2)}(L)\Big)
\ =\ \prod_{i=1}^{l_R} f^{(2)}(R_i)
\label{n2fact}
\ee
It is actually equal to  \cite{MMkon}
\be
{\rm coeff}_{Q^{(2)}_{R}\{\delta_{k,3}\}Q^{(2)}_{R}\{p_k\}}\Big(Z^{(2)}(L)\Big)
\ =\
{1\over 2^{5|R|/3}}\cdot\frac{Q_{2R}^{(2)}\{\delta_{k,3}\}}{Q_{R}^{(2)}\{\delta_{k,3}\}}
\frac{Q_{R}^{(2)}\{\delta_{k,1}\}}{Q_{2R}^{(2)}\{\delta_{k,1}\}}
\ee
which has exactly such factorization property
due to elegant factorization identities, see \cite{MMNO} for details. Here $2R$ means the Young diagram obtained from $R$ by doubling its line lengths.

Thus, one finally obtains
\be
Z^{(2)}(L)=\sum_{R\in S_2P}{1\over 2^{5|R|/3}}\cdot
{Q_R\{\delta_{k,1}\}\over Q_{2R}\{\delta_{k,1}\}}\cdot Q_R\{p_k\}Q_{2R}\{\delta_{k,3}\}
\ee

\subsection{The basic example: $n=3$}

After reminding the already known situation at $n=2$, we now make the first step into
{\it terra incognita} at $n>2$.
For $n+1=4$
\be
\left< \Tr Y^4\right> = \left<Y_{ij} Y_{jk} Y_{kl} Y_{li}\right>
=  2P_{ij,jk}P_{kl,li} + P_{ij,kl}P_{jk,li}
= \nn \\
= \frac{2\delta_{ik}}{(\lambda^2_i+\lambda_i\lambda_j+\lambda_j^2)(\lambda_k^2+\lambda_k\lambda_l+\lambda_l^2)}
+ \frac{\delta_{i,j,k,l}}{(\lambda^2_i+\lambda_i\lambda_j+\lambda_j^2)(\lambda_j^2+\lambda_j\lambda_k+\lambda_k^2)}
= \nn \\
= \sum_{ijl}\frac{2 }{(\lambda^2_i+\lambda_i\lambda_j+\lambda_j^2)(\lambda_i^2+\lambda_i\lambda_l+\lambda_l^2)}
+ \sum_i \frac{1}{9\lambda_i^4}
\ee
This correlator is {\it not} expressed through the time variables. However, there is another term of the same
{\it grading}, i.e. of the same degree in $L^{-1}$, with two extra powers of $L$ in the operator
compensated by those in the extra propagator: \
$\left<\Big(\Tr L Y^3\Big)^2\right>$.
When it is added to $\left<\Tr Y^4\right>$
with an appropriate coefficient, the sum gets expressed through the time variables:
\be
-\frac{1}{4}\left(\Big< \Tr Y^4\Big> +  2\left<\Big(\Tr L Y^3\Big)^2\right>\right) \
=\ {p_4+6p_1^2p_2 \over 36}
=
\label{n3first}
\ee
$$
=  {1\over 36}\left(\beta_1^{-1/2}\Big(7Q^{(3)}_{[4]}-\sqrt{3}Q^{(3)}_{[2,1,1]}\Big)
 -(2\sqrt{3}+i)  \Big(Q^{(3)}_{[2,2]} -iQ^{(3)}_{[3,1]}\Big)\right)
={\beta_2\over  27}
\sum_{\stackrel{R\in S_3P}{|R|=4\ }}c_R\cdot Q^{(3)}_R\{\delta_{k,4}\} Q^{(3)}_{R}\{p_k\}
$$
\vspace{-.5cm}

\noindent
with
\be
c_{[4]}=7,\ \ \ \ \ \ c_{[2,2]}=c_{[3,1]}=1-2i\sqrt{3},\ \ \ \ \ \ c_{[2,1,1]}=-1
\label{n3firstC}
\ee
Note that the diagram $[1,1,1,1]$, which does not belong to $S_3P$, is indeed missing at the r.h.s.

Similarly, in the next order
\be
\frac{1}{2!\cdot 4^2}\,\left(
\left< \Big(\Tr Y^4\Big)^2\right>-4\left<\Big(\Tr L Y^3\Big)^2\cdot\Tr Y^4\right>+{4\over 3}\left<\Big(\Tr L Y^3\Big)^4\right>\right)
=\nn\\
={1\over 32\cdot 81}\cdot\left(96p_7p_1+96p_5p_1^3+13p_4^2+156p_4p_2p_1^2-12p_2^4+ 36p_2^2p_1^4\right)
\label{n3second}
\ee
{\footnotesize
$$
\hspace{-.7cm} ={1\over 32\cdot 81}\left({5\cdot 7\cdot 11\over \sqrt{\beta_1}}Q^{(3)}_{[8]}+
5\cdot 7(1+4i\sqrt{3})Q^{(3)}_{[7,1]}-{5\cdot 7\over 2}(3\sqrt{3}i+13)Q^{(3)}_{[6,2]}-
5\beta_2\sqrt{3\beta_1}Q^{(3)}_{[6,1,1]}+{5\cdot 7\over 2}(11-5\sqrt{3}i)Q^{(3)}_{[5,3]}+3\cdot 5\sqrt{\beta_1}Q^{(3)}_{[5,2,1]}+
\right.
$$
$$
\hspace{-.7cm} \left.+{5\cdot 7\cdot 11\over 3\sqrt{3}}\beta_2^2Q^{(3)}_{[4,4]}+{5\sqrt{\beta_1}\over 2\sqrt{3}}
(23i+11\sqrt{3})\Big(Q^{(3)}_{[4,3,1]}-iQ^{(3)}_{[4,2,2]}\Big)-{2\cdot 7\over\sqrt{3}}\beta_2Q^{(3)}_{[4,2,1,1]}-{5\over\sqrt{3}}\beta_2^2\sqrt{\beta_1}Q^{(3)}_{[3,3,2]}
-{2\cdot 13\over 3}\beta_1^2\Big(Q^{(3)}_{[3,3,1,1]}+iQ^{(3)}_{[3,2,2,1]}\Big)
\right)
= 
$$
}
\be
=\left({\beta_2\over 27}\right)^2
\sum_{\stackrel{R\in S_3P}{|R|=8\ }}c_R\cdot Q^{(3)}_R\{\delta_{k,4}\} Q^{(3)}_{R}\{p_k\}
\ee
\vspace{-0.7cm}

\noindent
with
\be
c_{[8]}=c_{[4,4]}=5\cdot 7\cdot 11 ,\ \ \ \ \ \ c_{[7,1]}=c_{[6,2]}=c_{[5,3]}=-5\cdot 7\cdot(1+4i\sqrt{3}),\ \ \ \ \ \ c_{[6,1,1]}=c_{[5,2,1]}=c_{[3,3,2]}=-15,
\nn\\
\ \ \ \ \ \ c_{[4,3,1]}=c_{[4,2,2]}=-\frac{5(37-8i\sqrt{3})}{7},\ \ \ \ \ \ c_{[4,2,1,1]}=-7,
\ \ \ \ \ \ c_{[3,3,1,1]}=c_{[3,2,2,1]}=13 \ \ \ \ \ \ \ \ \ \ \
\ee

\bigskip

At $n=3$, one should not expect relations like (\ref{n2fact}) already because
they do not respect the selection rule for partitions from $S_3P$:
say, $[2,2]\in S_3P$, but $[2,2,2,2]\notin S_3P$.
Still, one can observe some interesting relations,
which resemble the corollaries of (\ref{n2fact}):
\be
c_{[3,3,1,1]}=c_{[3,2,2,1]}= |c_{[2,2]}|^2= |c_{[3,1]}|^2 \nn \\
c_{[4,2,1,1]} = c_{[4]}\cdot c_{[2,1,1]} \nn \\
\frac{c_{[4,3,1]}}{c_{[4]}c_{[3,1]}} = \frac{c_{[4,2,2]}}{c_{[4]}c_{[2,2]}}
\ee
however, say,
\be
c_{[4,4]}\neq c_{[4]}^2
\ee

We see from the above formulas that
extracting $Q^{(3)}\{\delta_{k,4}\}$ from the coefficients simplify them a little,
but the remaining pieces do not have any nice enough properties, e.g. do not factorize
in the spirit of \cite{Alex,MMNO},
as they did for $n=2$.
Worse than that, already in the next order, some of $Q^{(3)}_R\{\delta_{k,4}\}=0$,
though the corresponding contribution from the diagram $R$ is non-vanishing:
hence such an extraction is simply impossible in general situation.
The first diagrams with this property for $n=3$ appear at the level twelve:
$[ 5,5,1,1]$, $[5,3,2,1,1]$   and $[6,2,2,1,1]$.

In more detail,
\be
\frac{1}{3!\cdot 4^3}\,\left(
\left< \Big(\Tr Y^4\Big)^3\right>-6\left<\Big(\Tr L Y^3\Big)^2\cdot\Big(\Tr Y^4\Big)^2\right>
+4\left<\Big(\Tr L Y^3\Big)^4\cdot\Tr Y^4\right> - {8\over 15}\left<\Big(\Tr L Y^3\Big)^6\right>\right)
=\nn\\
= -\frac{5p_8p_2^2}{324}+\frac{5p_8p_1^4}{324}+\frac{7p_{10}p_1^2}{162} +\frac{25p_7p_4p_1}{972}
+\frac{5p_7p_2p_1^3}{162}
-\frac{p_5^2p_2}{162} +\frac{25p_5p_4p_1^3}{972}-\frac{p_5p_2^3p_1}{81} +\frac{p_5p_2p_1^5}{162}
+\frac{325p_4^3}{279936} -
\nn\\
 -\frac{25 p_4 p_2^4}{7776}
+\frac{25p_4p_2^2p_1^4}{2592}
-\frac{p_2^5p_1^2}{1296}+\frac{325p_4^2p_2p_1^2}{15552}  +\frac{p_2^3p_1^6}{1296}
\ \ = \ \ \left({\beta_2\over 27}\right)^3\!\!\!\!
\sum_{\stackrel{R\in S_3P}{|R|=12\ }} C_R\cdot Q^{(3)}_{R}\{p_k\}
\label{n3third}
\ee
Wherever possible, we present the much simpler and more ``symmetric" expressions for $c_R$
defined from $C_R = c_R\cdot Q^{(3)}_R\{\delta_{k,4}\}$
\be
c_{[12]}=c_{[8,4]} = 5\cdot 7\cdot 11\cdot 103, \ \ \ \ \ \ \ \ \ \ \ \nn\\
c_{[11,1]} = c_{[10,2]} = c_{[9,3]} = -5\cdot 7\cdot 11\cdot(59+54i\sqrt{3}), \ \ \ \ \ \ \
c_{[7,5]} = c_{[6,6]} = 5(2423-36\cdot 16 i\sqrt{3}), \ \ \ \
\nn \\
c_{[10, 1, 1]}=c_{[9, 2, 1]}= c_{[3,3, 2, 2, 1,1]}=175, \ \ \ \ \ \
c_{[8,3,1]}=c_{[8,2, 2 ]}=175(3i\sqrt{3}-8), \ \ \ \
c_{[7,4,1]}=35(15i\sqrt{3}-58), \ \ \ \
\nn \\
c_{[7,3,2]}=-5(187+6i\sqrt{3}), \ \ \ \
c_{[6, 5, 1]}=-5(355-78i\sqrt{3}), \ \ \ \
c_{[6, 4, 2]}=-5\cdot\frac{2233-582i\sqrt{3}}{19} \ \ \ \
\nn\\
c_{[6,3, 3 ]}=-5\cdot\frac{1345+354i\sqrt{3}}{7}, \ \ \ \
c_{[5, 5,2]}=25\cdot\frac{25 -12i\sqrt{3}}{7}, \ \ \ \
c_{[5, 4, 3]]}=-25\cdot\frac{23+132i\sqrt{3}}{7}, \ \ \ \
 \nn\\
 c_{[8,2,1,1]} = c_{[4,4,2,1,1]}=-385, \ \ \ \
c_{[7,3,1,1]}=c_{[7,2,2,1]}=35(23-6i\sqrt{3}), \ \ \ \
c_{[6,4,1,1]}=c_{[5,4,2,1]}=5(333-10i\sqrt{3}), \ \ \ \
\nn \\
c_{[6,3,2,1 ]}=5(179-80i\sqrt{3}), \ \ \ \
c_{[5,3,2, 2 ]}=c_{[5,3,3,1]}=5(113-70i\sqrt{3}), \ \ \ \
c_{[4,4,3,1]}= c_{[4,4,2,2]}=5\frac{323-466i\sqrt{3}}{7},\ \ \ \
\nn
\ee
\vspace{-0.5cm}
\be
c_{[4,3,3,2]}=-105, \ \ \ \ \ \ \ \
c_{[4,3,3,1,1]}=c_{[4,3,2,2,1]}=5(17+18i\sqrt{3}) \ \ \ \ \ \
\ee
but in the above mentioned cases, when
$Q^{(3)}_R\{\delta_{k,4}\}=0$ for $R=[5,5,1,1 ],\  [5,3,2,1,1 ], \ [6,2,2,1,1]$,
only $C_R$'s make sense:
\be
C_{[5,5,1,1]}    =     -45\cdot\frac{5+59i\sqrt{3}}{64}  \nn \\
C_{[5,3,2,1,1]}=    3^{1/4}\sqrt{2}\cdot 15\cdot\frac{19\sqrt{3}(1+i)-9(1-i)}{128} \nn \\
C_{[6,2,2,1,1]} =      3^{1/4}\sqrt{2}\cdot 15\cdot\frac{9(1+i) +19\sqrt{3}(1-i)}{128}
\ee

\section{Virasoro algebra action \label{VW}}

The generators of the positive part ($m>0$) of Virasoro algebra are
\be
\hat L_m^{(n)}:=\sum_{k=1}(k+nm)p_k{\partial\over\partial p_{k+nm}}
+{1\over 2}\sum_{k=1}^{nm-1}k(nm-k){\partial^2\over\partial p_k\partial p_{nm-k}}
\label{Vir}
\ee
It acts on the linear space of Schur functions, moreover, it leaves
the sub-space $S_nP$ intact, so that
\be
\hat L^{(n)}_m Q^{(n)}_R\{p\} =
\sum_{R',\ |R'|=|R|-mn} \xi^{(n,m)}_{R,R'}Q_{R'}^{(n)}\{p\}
\ee
e.g.
\be
\hat L^{(n)}_m Q^{(n)}_{[r]} = rQ^{(n)}_{[r-mn]}
\label{viracr}
\ee

\subsection{$n=2$}

For $n=2$ its action is known on $Q^{(2)}$ with time-variables, rescaled by $\sqrt{2}$ \cite{VirJ,MMNO}:
\be
\hat L_m^{(2)} Q^{(2)}_R\left\{\frac{p_k}{\sqrt{2}}\right\}
= \sum_{i=1}^{l_R} \frac{ (-)^{\nu_i}(R_i- m)}{(\sqrt{2})^{\delta_{R_i,m}}}
\cdot Q^{(2)}_{R-2m\epsilon_i}\left\{\frac{p_k}{\sqrt{2}}\right\}
\label{Virac2resc}
\ee
where $R-2m\epsilon_i$ means that exactly $i$-th length is diminished: $R_i\longrightarrow R_i-2m$.
This can make it shorter than some other lines
and thus imply reordering of lines in the diagram to put them back into decreasing order,
then $\nu_i(R,m)$ is the number of lines, which the $i$-th one needs to jump over,
e.g. $\hat L_2^{(2)} Q^{(2)}_{[6,5,3]}\left\{\frac{p_k}{\sqrt{2}}\right\}
= (6-2)Q^{(2)}_{[5,3,2]}\left\{\frac{p_k}{\sqrt{2}}\right\}
- (5-2)Q^{(2)}_{[6,3,1]}\left\{\frac{p_k}{\sqrt{2}}\right\}$
and
$\hat L_3^{(2)} Q^{(2)}_{[7,6,3]}\left\{\frac{p_k}{\sqrt{2}}\right\}
= (7-3)Q^{(2)}_{[6,3,1]}\left\{\frac{p_k}{\sqrt{2}}\right\}
- \frac{(6-3)}{\sqrt{2}} Q^{(2)}_{[7,3]}\left\{\frac{p_k}{\sqrt{2}}\right\}$.
If $R_i-2m=0$, then the line is simply omitted and the coefficient $1/\sqrt{2}$ appears.

Formula (\ref{Virac2resc}) looks reasonably nice, but expansion of partition functions
$Z^{(2)}_{GKM}$ in the basis $Q^{(2)}_R\left\{\frac{p_k}{\sqrt{2}}\right\}$  is rather ugly.
Expansion is nice in terms of $Q^{(2)}$ per se,
instead the Virasoro action on $Q^{(2)}$ per se is slightly more involved than (\ref{Virac2resc}):
\be
\hat L^{(2)}_m Q^{(2)}_{[r]} = rQ^{(2)}_{[r-2m]} \nn \\
\hat L^{(2)}_m Q^{(2)}_{[r,1]} =  rQ^{(2)}_{[r-2m,1]} + \sqrt{2}Q^{(2)}_{[r+1-2m]} \nn \\
\hat L^{(2)}_m Q^{(2)}_{[r,2]} =  rQ^{(2)}_{[r-2m,2]} + 2Q^{(2)}_{[r+1-2m,1]}, \ \ \ \ \ m\geq 2 \nn \\
\ldots
\ee

\subsection{Generic $n$}

For generic $n$, we note that the Virasoro algebra acts in the simplest way to slightly renormalized functions ${\cal Q}_{R}^{(n)}= \beta_1^{l_{_R}/2} Q_{R}^{(n)}$. Now the action of the operator $\hat L_1^{(n)}$ on ${\cal Q}_{R}^{(n)}$ is expanded into the $Q$ Schur functions at level $|R|-n$. The rule is that  $\hat L_1^{(n)}{\cal Q}_{R}^{(n)}$ spans only by Young diagrams $\check R=R-k_i\epsilon_i-k_j\epsilon_j$, $k_i+k_j=n$, and one of $k_i$ can be zero. This means that, for instance, in the case of $n=3$, there can be only either diagrams $R-3\epsilon_i$, or $R-2\epsilon_i-\epsilon_j$. Suppose that $\check R$ do not requite re-ordering the lines (i.e. the decreasing order is still preserved), and, moreover, the lines in all diagrams have different lengths (i.e. they are strict partitions). Then,
\be\label{L3}
\hat L_1^{(3)}{\cal Q}_{R}^{(3)}=\sum_i R_i{\cal Q}_{R-3\epsilon_i}^{(3)}+
\beta_2^{(3)}\sum_{i>j}{\cal Q}_{R-2\epsilon_i-\epsilon_j}^{(3)}+
\beta_1^{(3)}\sum_{i>j}{\cal Q}_{R-\epsilon_i-2\epsilon_j}^{(3)}
\ee
Similarly, in the case of $n=4$, there are possibilities: $R-4\epsilon_i$, $R-3\epsilon_i-\epsilon_j$ and  $R-2\epsilon_i-2\epsilon_j$ so that
\be
\hat L_1^{(4)}{\cal Q}_{R}^{(4)}=\sum_i R_i{\cal Q}_{R-4\epsilon_i}^{(4)}+
\beta_3^{(4)}\sum_{i>j}{\cal Q}_{R-3\epsilon_i-\epsilon_j}^{(4)}+
\beta_2^{(4)}\sum_{i> j}{\cal Q}_{R-2\epsilon_i-2\epsilon_j}^{(4)}+
\beta_1^{(4)}\sum_{i>j}{\cal Q}_{R-\epsilon_i-3\epsilon_j}^{(4)}
\ee
and generally
\be
\hat L_1^{(n)}{\cal Q}_{R}^{(n)}=\sum_i R_i{\cal Q}_{R-n\epsilon_i}^{(n)}+
\sum_{k=1}^{n-1}\beta_k^{(n)}\sum_{i>j}{\cal Q}_{R-k\epsilon_i-(n-k)\epsilon_j}^{(4)}
\ee
Moreover, this action is immediately continued to action of the general $\hat L_m^{(n)}$. For instance, instead of (\ref{L3}), one has now
\be
\hat L_m^{(3)}{\cal Q}_{R}^{(3)}=\sum_i R_i{\cal Q}_{R-3m\epsilon_i}^{(3)}+
\beta_2^{(3)}\sum_{i>j}{\cal Q}_{R-2m\epsilon_i-m\epsilon_j}^{(3)}+
\beta_1^{(3)}\sum_{i<j}{\cal Q}_{R-2m\epsilon_i-m\epsilon_j}^{(3)}
\ee
and similarly for other $n$: the coefficients do not depend on $m$:
\be
\hat L_m^{(n)}{\cal Q}_{R}^{(n)}=\sum_i R_i{\cal Q}_{R-nm\epsilon_i}^{(n)}+
\sum_{k=1}^{n-1}\beta_k^{(n)}\sum_{i>j}{\cal Q}_{R-km\epsilon_i-(n-k)m\epsilon_j}^{(4)}
\ee
Thus, ones gets
\be\label{Lf}
\hat L_m^{(n)}{\cal Q}_{[r]}^{(n)}=r{\cal Q}_{[r-nm]}^{(n)}\nn\\
\hat L_m^{(n)}{\cal Q}_{[r,k]}^{(n)}=r{\cal Q}_{[r-nm,k]}^{(n)}+\sum_{i=1}^{n-1}\beta^{(n)}_{n-i}{\cal Q}_{[r-nm+mi,k-mi]}^{(n)}+
k{\cal Q}_{[r,k-mn]}^{(n)}\ \ \ \ \ \ \
\hbox{for }r> nm+k\nn\\
\ldots
\ee

When the diagram have two lines of the same length, it acquires an additional factor of $\rho_1:=\sqrt{\beta_2^{(n)}/\beta_1^{(n)}}$. For instance,
\be
\hat L_m^{(n)}{\cal Q}_{[nm+k,k]}^{(n)}=r\rho_1 {\cal Q}_{[k,k]}^{(n)}+
\sum_{i=1}^{n-1}\beta^{(n)}_{n-i}{\cal Q}_{[k+ni,k-ni]}^{(n)}+k{\cal Q}_{[k+mn,k-mn]}^{(n)}
\ee
Similarly, for three lines of the same length, there is a factor of $\sim\sqrt{\beta_3^{(n)}}$, etc.

At last, when the diagram at the r.h.s. of (\ref{Lf}) requires a re-ordering of lines, each permutation needs an additional factor of $\rho_2:=1-\beta_1^{(n)}$. For instance,
\be
\hat L_1^{(4)}{\cal Q}_{[7,4,1]}^{(4)}=(\beta_3^{(4)}+7\rho_2){\cal Q}_{[4,3,1]}^{(4)}
+\beta_3^{(4)}\rho_1{\cal Q}_{[4,4]}^{(4)}+\beta_2^{(4)}{\cal Q}_{[5,2,1]}^{(4)}
+\beta_1^{(4)}\rho_1{\cal Q}_{[6,1,1]}^{(4)}+(\beta_3^{(4)}+4\rho_2){\cal Q}_{[7,1]}^{(4)}
\ee

Note also that there is an exception to the rule that only two lines are made shorter at the r.h.s. of these expressions: there could emerge a diagram $R-\sum_i^lk_i\epsilon_i$ with $l>2$ if two lines at the r.h.s. become of zero length. For instance,
\be
\hat L_1^{(n)}{\cal Q}_{[n+1,2,1]}^{(n)}=\Big((n+1)\rho_1\rho_2+\beta_{n-1}^{(n)}\rho_1\Big){\cal Q}_{[2,1,1]}^{(n)}+\beta_{n-1}^{(n)}\rho_1{\cal Q}_{[2,2]}^{(n)}+\beta_{n-2}^{(n)}\rho_2{\cal Q}_{[3,1]}^{(n)}+\underline{\beta_{n-1}^2{\cal Q}_{[4]}^{(n)}}
\ee
where the underlined term comes from $R-\epsilon_1-2\epsilon_2-\epsilon_3$.

\subsection{Virasoro constraints for GKM}

The partition function of GKM is annihilated \cite{GKM,MMM91,GN,Mikh} by action
of the Virasoro algebra \cite{FKN}
\be\label{Vcon}
\left(\hat L^{(n)}_m - \gamma_n (n+1+nm) \frac{\p}{\p p_{n+1+mn}}+{n^2-1\over 24}\delta_{m,0}+
{\delta_{m,-1}\over 2}\sum_{k=1}^{n-1}p_kp_{n-k}\right)\sum_R c_R^{(n)}Q_R^{(n)}\{p\} = 0,
\ \ \ \ \ \ m\ge -1
\ee
Here $\gamma_n$ is a constant that can be made arbitrary by a rescaling of the external matrix $L$ in (\ref{GKMint}). This results in a simple factor of $(const)^{|R|}$ in the summand in this formula. Our choice in this paper corresponds to $\gamma_n=n$.

Note that the first two constraints are rather trivial: both $\hat L^{(n)}_0$, $\hat L^{(n)}_{-1}$ gives rise to linear partial differential equations, and can be explicitly solved (see, e.g., \cite{AMMP}).

At $m>0$ one obtains
\be
\sum_{R',\ |R'|=|R|+mn} \xi_{R',R}^{(n,m)}c_{R'}^{(n)}
\ \ = \! \! \sum_{R'',\ |R''|=|R|+mn+n+1}  n\zeta_{R'',R}^{(n,n+1+mn)}c_{R''}^{(n)}
\label{virconmat}
\ee
where $\zeta$ is the matrix describing action of the derivative
$$
r\frac{\p}{\p p_{r}} Q_R^{(n)}\{p\} =
\sum_{R'\in S_{n}P(\nu)} \zeta^{(n,r)}_{R,R'}Q_{R'}^{(n)}\{p\}
$$
that is,
\be
r\frac{\p}{\p p_{r}} Q_R^{(n)}\{p\} =
\sum_{R'\in  S_nP(r)}\Psi^{(n)}_{R'}([r])\cdot Q_{R/R'}^{(n)}\{p_k\}
=\nn\\
 =\sum_{{R'\in S_n P(\nu)}\atop{\Delta\in O_n P(\nu)}} {\Psi^{(n)}_R(\Delta+r)\Psi^{(n)}_{R'}(\Delta)\over
\beta_\Delta z_\Delta}\cdot Q_{R'}^{(n)}\{p_k\}
 =\sum_{\Delta\in O_n P(\nu)}{\Psi^{(n)}_R(\Delta+r)\over z_\Delta}\cdot p_\Delta
\ee
where $\nu:=|R|-r$ and $\Delta+r$ denotes the Yang diagram with a line of length $r$ added.
E.g.
\be
r\frac{\p}{\p p_r} Q^{(n)}_{[l]} =  {\beta^{(n)}_r\over  \sqrt{\beta^{(n)}_1}}Q^{(n)}_{[l]/[r]}=\beta^{(n)}_r Q^{(n)}_{[l-r]}
\label{difacr}
\ee
since $\Psi^{(n)}_{[r]}([r])=\beta^{(n)}_r/\sqrt{\beta^{(n)}_1}$.
Thus,
\be
\zeta^{(n,r)}_{R,R'}=\sum_{\Delta\in O_n P(\nu)} {\Psi^{(n)}_R(\Delta+r)\Psi^{(n)}_{R'}(\Delta)\over
\beta_\Delta z_\Delta}
\ee
is not truly simple.

Note that $\hat L$ and $p$-derivative have different grading,
thus the Virasoro constraints relates coefficients $c_R$ with different sizes $|R|$.
For $m>0$, the Virasoro action is not injective, thus  (\ref{viracr}) and (\ref{difacr})
are not enough to check any constraint, one needs bigger pieces of
matrices $\xi$ and $\zeta$.
For $n=2$, the constraints with $m\geq -1$ are enough
to fix the partition function completely  (up to a common factor), while,
for $n>2$, one needs to add similar $W$-constraints
up to $W^{(n)}$, which can be studied in a similar way:
\be
\hat W^{(p|n)}_m\cdot\sum_R c_R^{(n)}Q_R^{(n)}\{p\} = 0,
\ \ \ \ \ \ m\ge 1-p,\ \ p=2,\ldots,n
\ee
Here $W^{(p|n)}$ denotes the $W$ algebra of spin $p$: $p=2$ corresponds to the Virasoro algebra, etc.

For instance, in the case of $n=3$, one has to add to the Virasoro constraints (\ref{Vcon}) the $W^{(3|3)}$-algebra constraints. This algebra at generic $n$ looks like \cite{FKN,GKM,Mikh}:
\be
\hat W^{(3|n)}_m:=\sum_{k=1}(k+l+nm)P_kP_l{\partial\over\partial p_{k+l+nm}}
+\sum_{k=1}\sum_{l=1}^{nm+k-1}l(nm+k-1)P_k{\partial^2\over\partial p_l\partial p_{nm+k-l}}+\nn\\
+{1\over 3}\sum_{k=1}^{mn-2}\sum_{l=1}^{mn-k-1}kl(mn-k-l){\partial^3\over\partial p_l\partial p_k\partial p_{mn-k-l}}+
{1\over 3}\sum_{k=1}^{-mn-2}\sum_{l=1}^{-mn-k-1}P_kP_lP_{-mn-k-l}
\ee
where $P_k:=p_k-n\delta_{n+1,k}$ at $k>0$, and $P_k=0$ otherwise.

\section{Conclusion}

To conclude,
we described an interesting set of functions $Q^{(n)}$,
which, in many respects, generalize the $Q$ Schur functions $Q^{(2)}$,
and provide a promising basis to expand the partition function of GKM
in the Kontsevich phase:
\be
\int \exp\left(\frac{\Tr X^{n+1}}{n+1} + \Tr L^NX\right) dX
\sim \sum_{R\in S_nP} C_R^{(n)} \cdot Q_R\{\Tr L^{-k}\}
\ee
This basis is distinguished by the selection rules:
independence of $p_{kn}$ and of diagrams beyond $S_nP$,
and, perhaps, also by integrability and Virasoro/W properties of $Q^{(n)}$,
which still need to be carefully formulated.

However, unlike the $n=2$ case, the coefficients $C_R^{(n)}$
are not properly identified and interpreted,
largely because of the calculation difficulties with $Q^{(n)}$ functions.
In the $n=2$ case, these coefficients have a very nice form \cite{MMkon}
$C_R^{(2)} = \frac{Q_{2R}\{\delta_{k,3}\}Q_{R}\{\delta_{k,1}\}}{Q_{2R}\{\delta_{k,1}\}}$
which has a profound combinatorial explanation \cite{Alex,MMNO},
and can be related to the superintegrability property of
the cubic Kontsevich model.
At the moment, it is an open problem if so defined {\it superintegrability}
persists for GKM at $n>2$.
We discuss this question
and a related issue of classification of correlators in the GKM
elsewhere.

\section*{Acknowledgements}

We are indebted to Sasha Alexandrov, John Harnad and Sasha Orlov
for numerous comments on \cite{MMkon},
which largely stimulated our further work in this direction.

This work was supported by the Russian Science Foundation (Grant No.20-12-00195).

\section*{Appendix}

In this Appendix, we illustrate our consideration by first terms of the $Q$-expansion of the GKM partition functions in the $n=4$ case.

The partition function of quintic GKM is defined as
\be
Z^{(4)} = \left<\exp\left(
-\frac{1}{5}\Tr Y^5  -  \Tr LY^4 - \Tr \Big(L^2 Y^3 +  LYLY^2\right)\right>_{\! 4}
:= \left<\exp\Big(-V_5  -  V_{1|4} - V_{2|3}\Big)\right>_{\! 4}
\ee
We introduced here a convenient notation $V_{a|b}$ for a term of the form $\Tr L^aX^b$.
Then the grading level, i.e. the power of an average $\prod_m V_{a_m|b_m}$ in $L^{-1}$,
is equal to $\sum_m\left(\frac{b_m}{2}-a_m\right)$.

The lowest grade terms in expansion of the partition function are:
\paragraph{Grading 5.}
\be
\left<-V_{1|4}+\frac{1}{2}V_{2|3}^2\right>_{\! 4} =
\left< - \Tr LY^4 +
\frac{1}{2}\left( \Tr \Big(L^2 Y^3 +  LYLY^2\Big)\right)^2\right>_{\! 4}
=\nn \\
= \frac{1}{32}\left(\frac{9}{ \sqrt{\beta_1^{(4)}}}Q_{[5]}^{(4)} -3(1-2i)Q_{[4,1]}^{(4)}-3(2+i)Q_{[3,2]}^{(4)}
-(1+4i)\sqrt{2}Q_{[3,1,1]}^{(4)}+(4-i)\sqrt{2}Q_{[2,2,1]}^{(4)} +\beta_3^{(4)}\sqrt{1-i}Q_{[2,1,1,1]}^{(4)}
\right)
= \nn
\ee
\be
= \frac{p_5+4p_3p_1^3+4p_2^2p_1}{32} \
= \ \frac{5\beta_3^{(4)}}{64}\sum_{\stackrel{R\in S_4P}{|R|=5\ }}c_R\cdot Q^{(4)}_R\{\delta_{k,5}\} Q^{(4)}_{R}\{p_k\}
\ \ \ \ \ \ \ \ \
\nn\\ \nn \\
c_{[5]}=9,\ \ \ \ \ \ c_{[4,1]}=c_{[3,2]}=3(1-2i),\ \ \ \ \ \ c_{[3,1,1]}=c_{[2,2,1]}=-(1+4i), \ \ \ \ \ \ \
c_{[2,1,1,1]}=-1 \ \ \ \ \ \ \ \ \ \
\ee
\paragraph{Grading 10.}
\be
\left<V_5V_{2|3}+ \frac{1}{2}V_{1|4}^2 -\frac{1}{2}V_{1|4}V_{2|3}^2+\frac{1}{24}V_{2|3}^4\right>_{\! 4}
= \left< \frac{1}{5} \Tr Y^5 \cdot \Tr \Big(L^2 Y^3 +  LYLY^2\Big)
+\frac{1}{2}\left( \Tr LY^4\right)^2 - \right.\nn \\  \left.
-\frac{1}{2} \Tr LY^4\left( \Tr \Big(L^2 Y^3 +  LYLY^2\Big)\right)^2
+\frac{1}{24}\left( \Tr \Big(L^2 Y^3 +  LYLY^2\Big)\right)^4
\right>_{\! 4}
= \left(\frac{5\beta_3^{(4)}}{64}\right)^2\sum_{\stackrel{R\in S_4P}{|R|=10\ }}c_R\cdot Q^{(4)}_R\{\delta_{k,5}\} Q^{(4)}_{R}\{p_k\}\nn
\ee
{\small
\be
c_{[10]} = c_{[5,5]} = 3^2\cdot 7^2, \ \ \ \ c_{[9,1]} = c_{[8,2]} = c_{[7,3]}=c_{[6,4]} = -9(3+52i),
\nn \\
c_{_{[8,1,1]}} = c_{_{[7,2,1]}} = c_{_{[6,3,1]}} = c_{_{[6,2,2]}} = -3(65+56i), \ \ \ \
c_{_{[5,4,1]}}=c_{_{[5,3,2]}} = -3(15+56i), \ \ \ \   c_{_{[4,4,2]}}=c_{_{[4,3,3]}} = \frac{3(11-148i)}{5},
\nn \\
c_{[7,1,1,1]} = c_{[6,2,1,1]} = -3, \ \ \ \  c_{[5,3,1,1]}=c_{[5,2,2,1]}= -\frac{3(47-14i)}{5}, \ \ \ \
c_{[4,4,1,1]} = -3(5-8i),
\nn \\
c_{[4,3,2,1]} = -3(3-8i), \ \ \ \ \ \ \ \
c_{[4,2,2,2]} = c_{[3,3,3,1]}=-\frac{3(9-32i)}{5}, \ \ \ \ \ \ c_{[3,3,2,2]} = 1,
\nn \\
c_{[3, 2, 2, 2, 1]} =3(7-6i), \ \ \ \ \ \ \ c_{[4, 2, 2, 1, 1]}= c_{[4, 3, 1, 1, 1]}= 3(7+6i),
\ \ \ \ \ \ \ c_{[3, 3, 2, 1, 1]} = 39,\ \ \ \ \ \ \  c_{[5, 2, 1, 1, 1]} = -9\nn
\ee
}
Similarly to the $n=3$ case, $Q^{(4)}_R\{\delta_{k,5}\}=0$ for $R=[3,2,2,1,1,1 ]$,
and only $C_R$ makes sense for this diagram:
\be
C_{[3,2,2,1,1,1]}=-{3^2\over 2^9}\sqrt{2}\beta_3^{(4)}
\ee
Note that $R=[3,2,2,1,1,1 ]$ is the only Young diagram out of $S_4P(10)$ that contains 6 lines.
\paragraph{Grading 15.}
The new phenomenon here is that the $L$-independent observable first contributes only to the
third ($k=3$) of the relevant levels $5k$:
\be
\left<\frac{1}{2}V_5^2 -V_5V_{1|4}V_{2|3} -\frac{1}{6}V_{1|4}^3
+\frac{1}{4}V_{1|4}^2V_{2|3}^2 + \frac{1}{6}V_5V_{2|3}^3
-\frac{1}{24}V_{1|4}V_{2|3}^4+\frac{1}{720}V_{2|3}^6
\right>_{\! 4}
\ee

\end{document}